# SOME REFINED RESULTS ON CONVERGENCE OF CURVELET TRANSFORM

Jouni Sampo
13.7.2012

Thesis of the author [1] stated that curvelet transform has non-linear approximation rate $M^{-2}$ for functions of two variable that are $C^5$ apart from $C^3$ edges. This means that extra smoothness allowed to remove log-factor from well known approximation rate $M^{-2}(\log(M))^3$ that hold if function is $C^2$ apart from $C^2$ edges. Here the some results from thesis are generalized by replacing $C^5$ smoothness assumption by $C^3$ assumption. For definitions, look [1].

**Theorem 1** *Let $f \in F_{N,n}$, $N \geq n$, $p \in S$ be point that minimizes $L = |D_a R_\theta (b-p)|$ and $\theta' \approx ka^{1/2}$ be the angle between major axis of $\gamma_{a\theta b}$ and tangent of $S$ at point $p$. Then for any $K > 0$ and $0 < \varepsilon < 2$*

$$\left| \int_{\mathbb{R}^2} f(x) \gamma_{a\theta b}(x) dx \right|$$
$$\lesssim \begin{cases} \max\left\{ a^{3/4+N}, \frac{a^{3/4}}{1+L^K} \right\} & , \quad |\theta'| \lesssim a^{1/2} \\ \max\left\{ a^{3/4+N}, \frac{a^{3/4}}{1+|k|^K L^K} \right\} & , \quad n \geq 3, \ |\theta'| \gtrsim a^{1/2}, \ |k|^{1-\varepsilon/2} \gtrsim L \\ \max\left\{ a^{3/4+N}, \frac{a^{3/4}}{1+|k|^{3+\varepsilon} L^K} \right\} & , \quad n \geq 3, \ |\theta'| \gtrsim a^{1/2}, \ |k|^{1-\varepsilon/2} \lesssim L \end{cases} \quad (1)$$

**Proof.** First we assume that $L \lesssim |k|^{1-\varepsilon/2}$. With this assumption the point $p$ is on major axis of $\gamma$ (see proof of theorem 15 in [1]).

Let assume that for regions $R_i \subset \mathbb{R}^2$ holds

$$\cup_{i=-2}^{3} R_i = \mathbb{R}^2 \quad \text{and} \quad i \neq j \quad \Rightarrow \quad R_i \cap R_j = \emptyset \quad (2)$$

and define
$$f_i = 1_{R_1} f. \quad (3)$$

Then
$$\int_{\mathbb{R}^2} \gamma_{a\theta b} f = \sum_{i=1}^{4} \int_{\mathbb{R}^2} \gamma_{a\theta b} f_i. \quad (4)$$

Next we assume that for regions $\tilde{R}_i \subset \mathbb{R}^2$ hold

$$R_i \cap \tilde{R}_i = \emptyset. \quad (5)$$

With any functions $P_i$ we can write

$$\begin{aligned}
&\int_{\mathbb{R}^2} \gamma_{a\theta b} f_i \\
&= \int_{\mathbb{R}^2} \gamma_{a\theta b}(1_{R_i} f - 1_{R_i} P_i + 1_{R_i} P_i) \\
&= \int_{\mathbb{R}^2} \gamma_{a\theta b}(1_{R_i} f - 1_{R_i} P_i) + \int_{\mathbb{R}^2} \gamma_{a\theta b} 1_{R_i} P_i \\
&= \int_{\mathbb{R}^2} \gamma_{a\theta b}(1_{R_i} f - 1_{R_i} P_i) + \int_{\mathbb{R}^2} \gamma_{a\theta b}(1_{R_i} P_i + 1_{\tilde{R}_i} P_i - 1_{\tilde{R}_i} P_i) \\
&= \int_{\mathbb{R}^2} \gamma_{a\theta b}(1_{R_i} f - 1_{R_i} P_i) + \int_{\mathbb{R}^2} \gamma_{a\theta b} 1_{R_i \cup \tilde{R}_i} P_i - \int_{\mathbb{R}^2} \gamma_{a\theta b} 1_{\tilde{R}_i} P_i) \\
&= \int_{R_i} \gamma_{a\theta b}(f - P_i) + \int_{R_i \cup \tilde{R}_i} \gamma_{a\theta b} P_i - \int_{\tilde{R}_i} \gamma_{a\theta b} P_i
\end{aligned} \quad (6)$$

Now, for $i = -2, \ldots, 2$, we define regions $R_i$ and $\tilde{R}_i$ as illustrated in Figure 1 and

$$R_3 = \setminus \cup_{i=-2}^{2} R_i. \quad (7)$$

Regions $R_0$, $R_1$ and $R_2$ are also illustrated with greater details in Figures 2, 3, 4. On those figures lengths $d$ and $h$ are as follows;

$$d \approx a^{1/2} k^{-1}, \quad h \approx a k^{-\varepsilon}. \quad (8)$$

Also $\theta'$ is assumed to be small, i.e. $\sin(\theta') \approx \theta'$. Case of "big" angles is omitted here since acceptable estimates are very straightforward to produce in that case (one can for example take $d \approx a$ and reproduce the rest of proof quite straight).

First, in the case $i = 3$, by rapid decay of $\gamma$, we get

$$\int_{\mathbb{R}^2} \gamma_{a\theta b} f_3 \lesssim \frac{a^{3/4}}{1 + k^{(1-\varepsilon/2)K} L^K}. \quad (9)$$

We concentrate next to case $i = 2$. If we define $P_2$ such that it is polynomial in direction of minor axis of $\gamma_{a\theta b}$, then, since vanishing moments of $\gamma_{a\theta b}$,

$$\int_{R_2 \cup \tilde{R}_2} \gamma_{a\theta b} P_2 = 0. \quad (10)$$

Moreover, because rapid decay of $\gamma$ and dimensions of $R_2$ (look figure 2),

$$\left| \int_{\tilde{R}_2} \gamma_{a\theta b} P_2 \right| \lesssim \frac{a^{3/4}}{1 + k^{(1-\varepsilon/2)K} L^K}. \quad (11)$$

By choosing $P_2$ so that slices of $P_2$ in the direction of minor axis of $\gamma_{a\theta b}$ are always second order Taylor polynomials, developed at major axis of $\gamma_{a\theta b}$ for slices of $f$, we get

$$\left| \int_{R_2} \gamma_{a\theta b} f \right| \lesssim a^{3/4+3}. \quad (12)$$

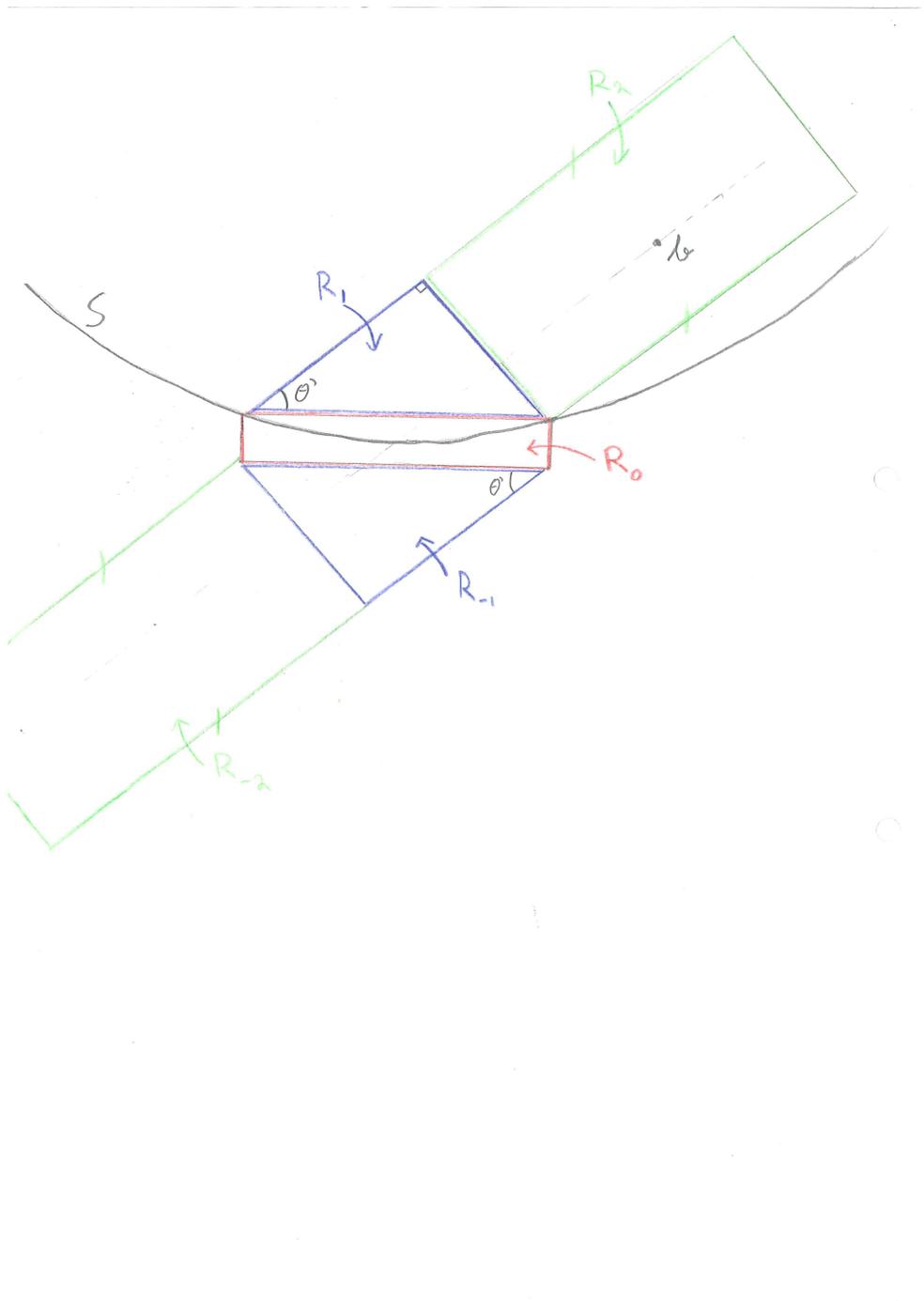

Figure 1: Illustration of regions $R_i$.

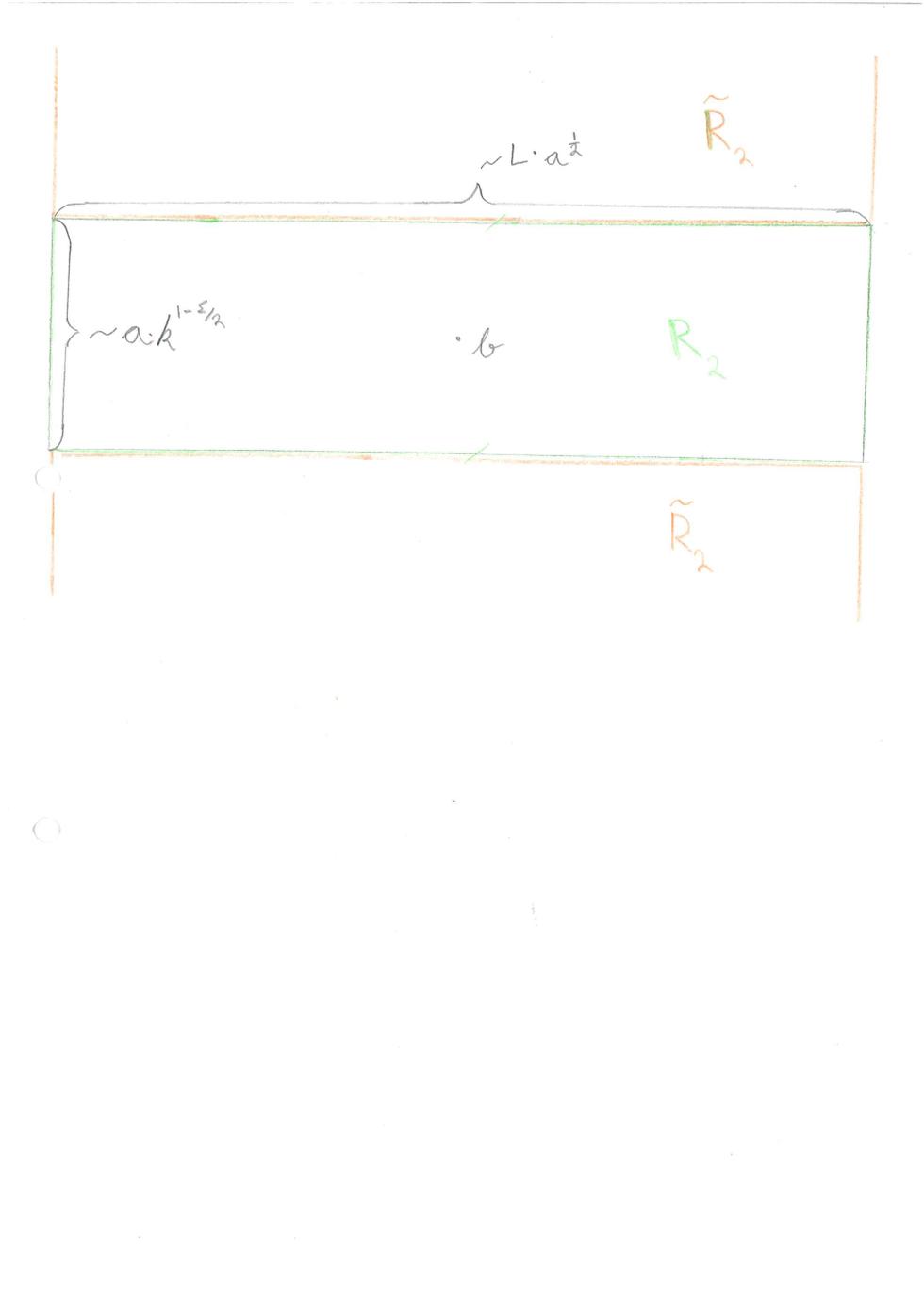

Figure 2: (a) Illustration of regions $R_2$ and $\tilde{R}_2$.

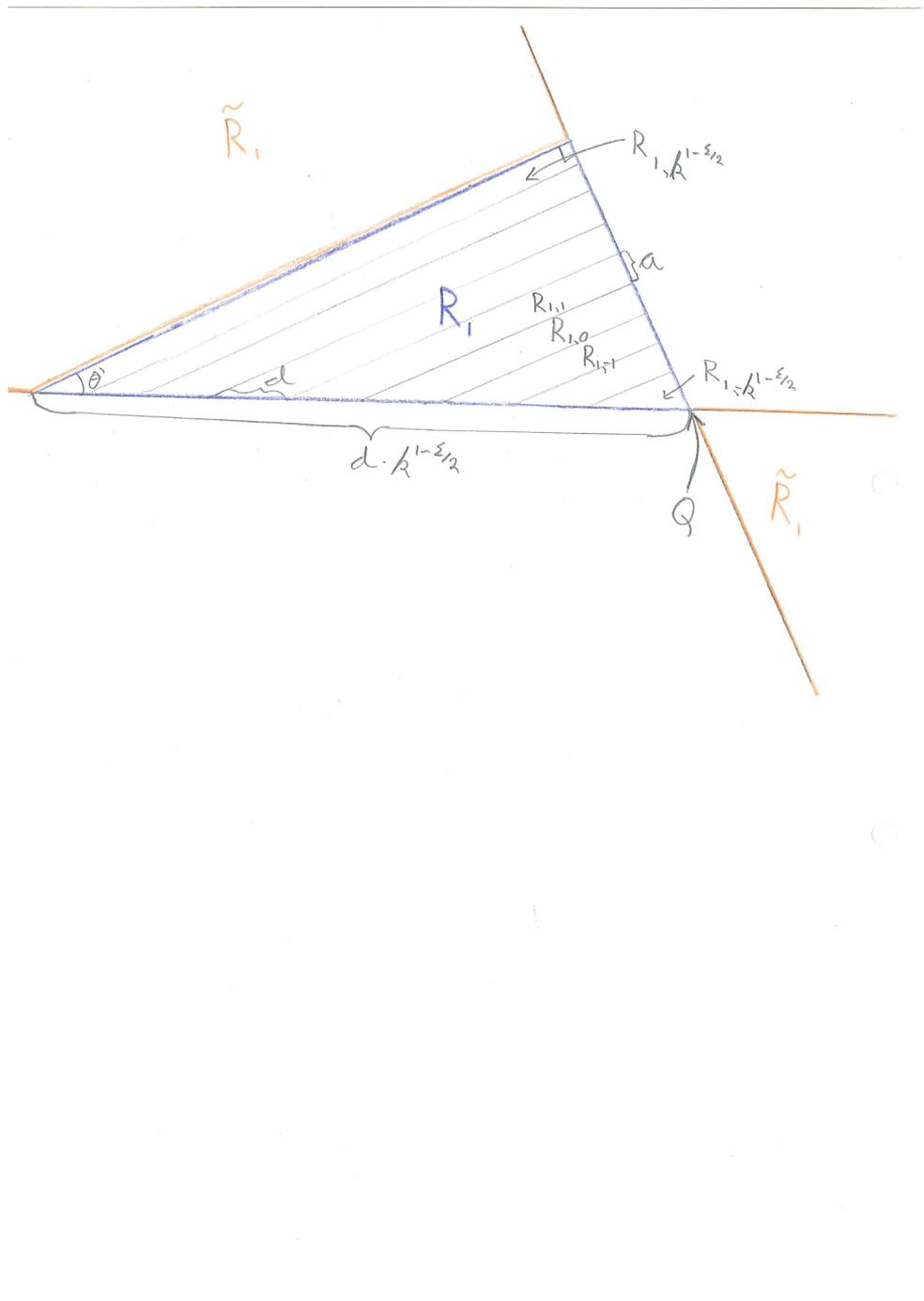

Figure 3: Illustration of regions $R_1$ and $\tilde{R}_1$.

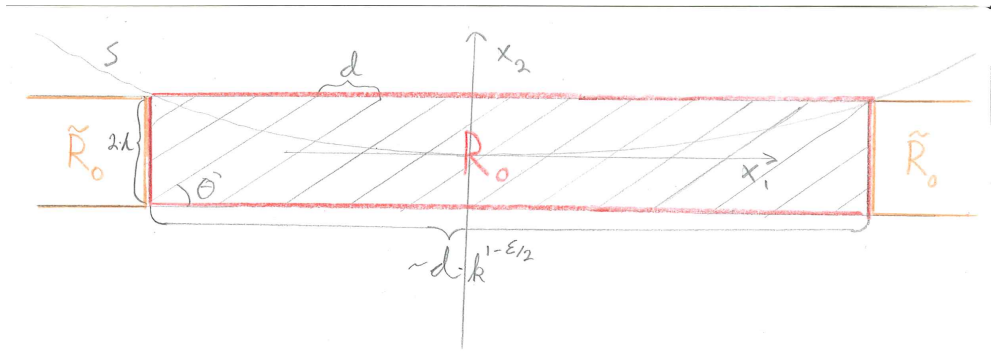

Figure 4: Illustration of regions $R_0$ and $\tilde{R}_0$.

Now we look the case $i = 1$. In this case we again define $P_1$ by using second order Taylor polynomials. However this time we define those along radial lines that intersect in point $Q$ in Figure 3. Points where polynomials are developed can be at major axis of $\gamma_{a\theta b}$.

By doing integration in the polar coordinates $r$ and $\alpha$ (origo at point $Q$) we get

$$\int_{R_1 \cup \tilde{R}_1} \gamma_{a\theta b} P_1 = 0. \tag{13}$$

This is because when when integrating first respect to $r$ we get zero (because of vanishing moments of $\gamma_{a\theta b}$.

By looking shape of $R_1$ and using rapid decay of $\gamma_{a\theta b}$ we get again

$$\left| \int_{\tilde{R}_1} \gamma_{a\theta b} P_1 \right| \lesssim \frac{a^{3/4}}{1 + |k|^{(1-\varepsilon/2)K} L^K}. \tag{14}$$

When making estimate for $\left| \int_{R_1} \gamma_{a\theta b}(f - P_1) \right|$, we divide $R_1$ into subregions $R_{1,l}$ like illustrated in figure XX. First we notice that because of the way how we defined $P_1$,

$$x \in R_{1,l} \quad \Rightarrow \quad |f(x) - P_1(x)| \lesssim (ld)^3. \tag{15}$$

Also area $m(R_{1,l})$ of region $R_{1,l}$ is bounded by

$$m(R_{1,l}) \lesssim m(R_{1,|k|^{1-\varepsilon/2}}) \approx as_2 \approx a^{3/2} k^{-\varepsilon/2}. \tag{16}$$

By using these two estimate and rapid decay of $\gamma$, we get

$$\begin{aligned}
&\left| \int_{R_1} \gamma_{a\theta b}(f - P_1) \right| \\
&\lesssim \sum_{l=-|k|^{1-\varepsilon/2}}^{|k|^{1-\varepsilon/2}} \int_{R_{1,l}} \frac{a^{-3/4}}{1 + |l|^K L^K} (ld)^3 \\
&\lesssim \frac{a^{3/2} k^{-\varepsilon/2} d^3}{1 + L^K} \\
&\approx \frac{a^{3/2 - 3/4 + 3/2} k^{-3-\varepsilon/2}}{1 + L^K}
\end{aligned} \tag{17}$$

This is clearly acceptable.

Cases $i = -1$ and $i = -2$ are identical to cases $i = 1$ and $i = 2$, so those are omitted.

Now only the case $i = 0$ is left. But $R_0$ is exactly the same region as $R_T$ in [1] and it can be handled similar techniques as in there. For the sake of completeness we give here the major steps (and also correct some minor missprints).

The major point is that $R_0$ is heavily "corrupted" by discontinuity curve $S$, i.e. we cant apply Taylor polynomials directly. Instead of that, we do such a change of variable that "twist" $S$ to straight line inside region $R_0$.

First we define $x_1$- and $x_2$-coordinate axes as illustrated in Figure 4. Inside $R_0$, with small enough scales $a$, curve $S$ can be considered as function $g(x_1)$.

We define the twisting operator $T_g : R_0 \to R_0$ by formula

$$T_g(x_1, x_2) := \begin{cases} (x_1, x_2 + \frac{h-|x_2|}{h}g(x_1)) &, \quad (x_1, x_2) \in R_0 \\ (x_1, x_2) &, \quad (x_1, x_2) \notin R_0 \end{cases} \tag{18}$$

Notice that
$$T_g R_0 = R_0 \tag{19}$$

and if apart from $S$, $f$ is three times continuously differentiable (with bounded derivatives) and first three derivatives of $g$ are continuous and bounded, then first 3 derivatives of the function
$$\tilde{f}(x) := f(T_g x), \tag{20}$$
in direction of $x_1$-axis are bounded and continuous inside $R_0$.

Change of variable
$$x = T_g y, \tag{21}$$

$$dx = \begin{cases} \det(J(y))dy &, \quad y \in R_0 \\ dy &, \quad y \notin R_0 \end{cases}$$
$$= \begin{cases} \frac{\partial x_2}{\partial y_2}dy &, \quad y \in R_0 \\ dy &, \quad y \notin R_0 \end{cases} \tag{22}$$
$$= \begin{cases} (1 + sgn(y_2)\frac{g(y_1)}{h})dy &, \quad y \in R_0 \\ dy &, \quad y \notin R_0 \end{cases},$$

gives
$$\left| \int_{R_0} f(x)\gamma_{a\theta b}(x)dx \right| = \left| \int_{R_0} \tilde{f}(y) \det(J(y))\tilde{\gamma}_{a\theta b}(y)dy \right|, \tag{23}$$

where
$$h(y) := \tilde{f}(y) \det(J(y)) = \begin{cases} \tilde{f}(y) \det(J(y)) &, \quad y \in R_0 \\ f(y) &, \quad y \notin R_0 \end{cases} \tag{24}$$

and
$$\tilde{\gamma}_{a\theta b}(y) := \gamma_{a\theta b}(T_g y). \tag{25}$$

Since
$$|g(y_1)| \lesssim y_1^2, \quad |g'(y_1)| \lesssim |y_1|, \quad 0 \leq \frac{h-|y_2|}{h} \leq 1, \tag{26}$$

it is quite clear that
$$|h(y)| \lesssim 1, \quad y \in \mathbb{R}^2, \tag{27}$$

$$\left| \frac{\partial h(y)}{\partial y_1} \right| \lesssim \begin{cases} \frac{|y_1|}{h} &, \quad y \in R_0 \\ 1 &, \quad y \notin R_0 \end{cases} \tag{28}$$

and for $2 \leq m \leq 3$

$$\left|\frac{\partial^m h(y)}{\partial y_1^m}\right| \lesssim \begin{cases} \frac{1}{h} &, \quad y \in R_0 \\ 1 &, \quad y \notin R_0 \end{cases} \quad (29)$$

On the border of $R_0$ the function $\tilde{\gamma}_{a\theta b}$ is discontinuous but all decay properties of $\tilde{\gamma}_{a\theta b}$ (and it's derivatives) remain. Also

$$\tilde{\gamma}(y) = \gamma(y), \forall y \notin R_0. \quad (30)$$

However, unlike $\gamma_{a\theta b}$, the function $\tilde{\gamma}_{a\theta b}$ does not have directional vanishing moments. Because of that we will "recreate" function $\gamma_{a\theta b}$:

$$\begin{aligned}
&\int_{R_0} f(x)\gamma_{a\theta b}(x)dx \\
&= \int_{R_0} h(y)\tilde{\gamma}_{a\theta b}(y)dy \\
&= \int_{R_0} h(y)(\tilde{\gamma}_{a\theta b}(y) - \gamma_{a\theta b}(y) + \gamma_{a\theta b}(y))dy \\
&= \int_{R_0} h(y)(\tilde{\gamma}_{a\theta b}(y) - \gamma_{a\theta b}(y))dy + \int_{R_0} h\gamma_{a\theta b}(y)dy.
\end{aligned} \quad (31)$$

The last of these two integral is handled now very similarly what we did with region $R_2$: we create polynomial $P_0$ by defining it as second order Taylor polynomials separately for each slice of function $h$ that are aligned to $x_1$-axis. This way we get

$$\int_{R_0} h\gamma_{a\theta b} = \int_{R_0} \gamma_{a\theta b}(h - P_0) + \int_{R_0 \cup \tilde{R}_0} \gamma_{a\theta b} P_0 - \int_{\tilde{R}_0} \gamma_{a\theta b} P_0. \quad (32)$$

Because of vanishing moments of $\gamma$ we have

$$\int_{R_0 \cup \tilde{R}_0} \gamma_{a\theta b} P_0 = 0, \quad (33)$$

and because of rapid decay of $\gamma$ we have

$$\left|\int_{\tilde{R}_0} \gamma_{a\theta b} P_0\right| \lesssim \frac{a^{3/4}}{1 + |k|^{(1+\varepsilon/2)K} L^K}. \quad (34)$$

By using regularity of function $h$, rapid decay of $\gamma$ and dimensions of $R_0$ we get with straightforward calculation that

$$\left|\int_{R_0} \gamma_{a\theta b}(h - P_0)\right| \lesssim \frac{a^2}{1 + |k|^4 L^K} a^{-3/4} \quad (35)$$

and by remembering also that $\gamma$ is differentiable (infinitely many times), we get

$$\left|\int_{R_0} h(y)(\tilde{\gamma}_{a\theta b}(y) - \gamma_{a\theta b}(y))dy\right| \leq \frac{a^{3/4}}{1 + |k|^{3+\varepsilon} L^K}. \quad (36)$$

Details for these calculations can be founded from proof of theorem 14 in [1] (only difference is that here we already write factor $L^K$, that follows from rapid decay of $\gamma$, visible already here).

Finally we turn to case $L \gtrsim k^{1-\varepsilon/2}$. Here we get straightforwardly estimate

$$\left| \int_{R_0} \gamma_{a\theta b} f \right| \lesssim \frac{a^{3/4+3}}{1+L^{2K}} \leq \max\{ \frac{a^{3/4}}{1+k^{(1-\varepsilon/2)K}L^K}, a^{3/4+3} \} \tag{37}$$

by using similar techniques as with case $i = 2$ before. Notice that all exponents related to $K$ can be simplified to form used in theorem since there is not any limitation for $K$ and $1 - \varepsilon/2 > 0$. ∎

**Theorem 2** *Let $f_{M,C}$ be $M$-term non-linear approximation of $f$ by using curvelets. If $f \in F_{3,3}$, then $\|f - f_{M,C}\|_2^2 \leq \mathcal{O}(M^{-2})$.*

**Proof.** Proof is exactly same as proof of Theorem 17 in [1], now we just can apply the improved decay estimate (1). ∎